\documentclass[aps,twocolumn]{revtex4}
\usepackage{amsmath,amssymb,amssymb,graphicx,color,bbm}
\usepackage[mathscr]{eucal}
\usepackage{epsfig}

\newcommand{\beq}{\begin{equation}}
\newcommand{\eeq}{\end{equation}}
\newcommand{\beqa}{\begin{eqnarray}}
\newcommand{\eeqa}{\end{eqnarray}}

\newcommand{\tr}{\mathop{\mathrm{tr}}}

\def\one{\ensuremath{\hbox{$\mathrm I$\kern-.6em$\mathrm 1$}}}
\def\tr{ \mbox{tr}}

\begin{document}

\title{Optimal Matrix Product States for the Heisenberg Spin Chain}

\author{  Jos\'e I. Latorre and Vicent Pic\'o}
\affiliation{Dept. Estructura i Constituents de la Mat\`eria, Universitat de Barcelona, 08028 Barcelona, Spain.}

\begin{abstract}
We present some exact results for the optimal Matrix Product State (MPS) approximation to the ground state of the infinite isotropic Heisenberg spin-1/2 chain. Our approach is based on the systematic use of Schmidt decompositions to reduce the problem of approximating for the ground state of a spin chain to an analytical minimization. This allows to show
that results of standard simulations, {\sl e.g.} density matrix renormalization group and infinite time evolving block decimation, do correspond to the result obtained by this minimization strategy and, thus, both methods deliver optimal MPS with the same energy but, otherwise, different properties. We also find
that translational and rotational symmetries cannot be maintained simultaneously by the MPS ansatz of minimum energy and present explicit constructions for each case. Furthermore,
we analyze symmetry restoration and quantify it to uncover new scaling relations. 
The method we propose can be extended to any translational invariant Hamiltonian.

\end{abstract}

\maketitle

%%%%%%%%%%%%%%%%%%%%%%%%%%%%%%%%%%%%%%%%%%%%%%%%%%%%%%%%%%%%%%%%%%%%%%%%%%%%%%
%%%%%%%%%%%%%%%%%%%%%%%%%%%%%%%%%%%%%%%%%%%%%%%%%%%%%%%%%%%%%%%%%%%%%%%%%%%%%%
\section{Introduction}
%%% Strongly correlated systems.
Numerical simulations stand as a fundamental tool to analyze strongly
correlated quantum systems. Basically, two different approaches have
proven very powerful. The first one corresponds to Monte Carlo simulation,
which can be applied to most systems with the exception of those
that present a sign problem. A second approach 
is based on introducing a structured ansatz for {\sl e.g.} the ground state of the 
system that captures as much of its entanglement properties as possible.
The kind of structure which is best suited for a given system will depend
on its dimensionality, symmetry properties and geometry of interactions. 

A widely used structured ansatz to handle one-dimensional systems corresponds to 
Matrix Product States (MPS) \cite{Ostlund:1995,Perez-Garcia:2007}. The basic idea of this approach
consists of representing states via their correlations rather than their
coefficients in  position or momentum space. Quantum states that carry little
entanglement are well described by MPS \cite{VLRK,Vidal:MPS}. This is the case of one-dimensional
off-critical states. There, MPS produces a very precise approximation to all
the properties of the state. The MPS approach deteriorates at criticality since
conformal invariance is recovered and quantum correlations pervade the system.
Yet, MPS remains a useful tool in that regime. In higher dimensions, MPS must be
substituted with the so-called Projected Entangled Pairs (PEPs) \cite{Cirac:PEPS}. 
A conceptually different structured ansatz corresponds to tensor networks
that do not encode the standard nearest neighbor topology but, rather, 
a renormalization group tree. Tree tensors and, in particular, the Multiscale Entanglement Renormalization
Ansatz (MERA) \cite{Vidal:MERA} may well turn into a powerful method to deal with critical quantum systems
in any number of dimensions . In this paper, we shall concentrate on
finding exact results for the MPS approximation.

The issue of finding an optimal MPS representation for the ground state
of an infinite quantum system is not trivial. 
A number of results can be found in the literature \cite{MPSexact} (and references therein)
concerning exact MPS in particular cases,  exploiting rotational
or translational invariance.
It is also possible, for instance,
to rewrite the Laughlin wave function coefficients as a product of matrices
\cite{ILO:2006}. But, in general, exact results are excepcional and
it is necessary to resort to
a specific numerical algorithm that delivers such a state. In the case
of infinite systems, two possibilities are
at hand: the Density Matrix Renormalization Group (DMRG) algorithm 
\cite{White:1992} and the Euclidean evolution (iTEBD) algorithm \cite{Vidal:iTEBD}. Both methods have been proven to produce the same results on examples \cite{TOIL:2008}. Nevertheless, it remains
an open question whether both algorithms do find the optimal MPS approximation
to the ground state of a given Hamiltonian. It is also unclear how the MPS approximation
handles the symmetries present in a Hamiltonian. 
We shall try to answer both of this questions in this paper.

Our approach will exploit the iTEBD approach in an analytical way and turn the
problem of finding optimal MPS to a minimization strategy.  
Thus, the novelty of the method lies in the exactness of this minimization procedure
that allows for an explicit analysis of symmetries in the MPS context.
We shall present our method and results for the particular case
of the isotropic spin-$1/2$ anti-ferromagnetic Heisenberg chain. 
This thoroughly studied model is suited for our purposes since the
MPS approximation will make a compromise  to handle its symmetries, 
as we shall see later on.

We divide our paper as follows. First, in Sect. \ref{Sect:Heisenberg},
we recall some basic exact results for the Heisenberg spin chain that we 
use as a testing ground for the MPS approach. Sect. \ref{Sect:iTEBD} is devoted
to present the elements of MPS states and the iTEBD algorithm.
Then, in Sect. \ref{Sect:exact}, we transform the iTEBD approach into
a minimization strategy and find exact solutions for MPS states. 
The analysis on symmetry restoration is discussed
in Sect. \ref{Sect:symmetry}.

%%%%%%%%%%%%%%%%%%%%%%%%%%%%%%%%%%%%%%%%%%%%%%%%%%%%%%%%%%%%%%%%%%%%%%%%%%%%%%
%%%%%%%%%%%%%%%%%%%%%%%%%%%%%%%%%%%%%%%%%%%%%%%%%%%%%%%%%%%%%%%%%%%%%%%%%%%%%%
\section{Heisenberg spin chain}
\label{Sect:Heisenberg}

The isotropic infinite spin-$1/2$ Heisenberg quantum chain is described 
by the Hamiltonian 
\begin{eqnarray}
  H &=& \sum_m h_{[m,m+1]} \nonumber \\ 
          &=&\frac{1}{4}\sum_{m} \left( \sigma_m^X \sigma_{m+1}^X 
            + \sigma_m^Y \sigma_{m+1}^Y + \sigma_m^Z \sigma_{m+1}^Z \right),
  \label{eq:ham}
\end{eqnarray}
where $\sigma_m^X$, $\sigma_m^Y$ and $\sigma_m^Z$ are 
Pauli matrices at site $m$. This theory stands
as a reference model
to try new ideas to treat strongly correlated
systems. There are a several reasons that justify
its relevance. First, the ground state of the Heisenberg 
spin chain displays critical
phenomena, that is, long distance correlations dominate the system and 
entanglement entropy of a finite block scales logarithmically with its size
\cite{VLRK}.
Second, the conformal properties of the Heisenberg chain
are related to the rotational symmetry of the Hamiltonian.  
This will be a key ingredient in our analysis. 
Third, the Heisenberg model is exactly solvable using the Bethe ansatz
{\cite{Bethe:1931,Sachdev:1999}}.

%Ground State properties
Let us be more precise about some of the properties of the Heisenberg chain that
can be calculated exactly. The exact energy of the ground state of this model can be worked out via the Bethe ansatz, 
$E = 1/4-\ln 2 \sim -0.443147181\dots$ .
The iTEBD algorithm that we shall present later on, will work as a variational method  based on energy minimization
strategy. Therefore, it should approach the above result from above.

%Density matrix. 2-body & 1-body
The ground state of the Heisenberg Hamiltonian is translationally invariant and,
as a consequence of rotational invariance, can be chosen to be an eigenstate of the total $Z$-component of spin, with eigenvalue equal to 0. These properties constrain the form of the state and, thus, its
reduced density matrices. Let us analyze this point in more detail.

We start with  an arbitrary state $\vert \psi\rangle$.
Its 2-body reduced density matrix for two adjacent sites,
where all spins are integrated out but for the sites $m$ and $m+1$, reads
\begin{equation}
\rho_{\left[m,m+1\right]}= \tr_{\left[m,m+1\right]}\vert\psi\rangle\langle\psi\vert \ .
\end{equation}
This operator takes the general form
\begin{equation}
\rho_{\left[m,m+1\right]} = \left( \begin{array} {cccc}
a & d & d' & f \\
d & b & c & e \\
d' & c & b' & e' \\
f & e & e' & a' \\
\end{array} \right) \ . \label{eq:twobody}
\end{equation}
We can now compute the 1-body reduced density matrix as a function of the 2-body density matrix parameters. Tracing out spin $m+1$ we get
\begin{equation}
\rho_{\left[m\right]} = \left( \begin{array} {cc}
a+b & d'+e \\
d'+e & b'+a' \\
\end{array} \right) \ . \label{eq:onebody}
\end{equation}

It is clear that
the total amount of parameters that describe the 1-body and 2-body
ground state reduced density matrices should be reduced by the presence of symmetries, as it is the case in the Heisenberg chain.
Let us consider separately how translational invariance and
rotational invariance simplify the above general discussion for
a finite spin chain.

\subsubsection{Eigenstate of the total $Z$-component of spin}
An eigenstate of the total $Z$-component of spin with eigenvalue $\Lambda$ obeys
\begin{equation}
\sum_{m=1}^N{\sigma_m^Z} { | \psi \rangle } = \Lambda | \psi \rangle \ . \label{eq:eigenstate}
\end{equation} 
It is possible to write such a state in the spin basis
\begin{equation}
\sum_{m=1}^N{\sigma_m^Z} | \psi \rangle = \sum_{i_1 \ldots i_N}^{\lbrace -1,1 \rbrace} { c_{i_1 \ldots i_N} \sum_{m=1}^N {\sigma_m^Z} { | i_1 \ldots i_N \rangle } } \ ,
\end{equation}
where the coefficients $c_{i_1 \ldots i_N}$ will be constrained. It is straightforward to see that, for a given $\Lambda$, 
\begin{equation}
\left ( \sum_{m=1}^N {i_m} \right ) c_{i_1 \ldots i_N}  = \Lambda c_{i_1 \ldots i_N}
\end{equation}
implies
\begin{equation}
\sum_{m=1}^N {i_m} \neq \Lambda \longrightarrow c_{i_1 \ldots i_N} = 0 \ . \label{eq:vanish_c}
\end{equation}
Furthermore, we can write the 2-body density matrix as a function of the basis elements
\begin{equation}
\rho_{\left[m,m+1\right]} = \rho^{i_m i_{m+1}}_{i'_m i'_{m+1}} | i_m i_{m+1} \rangle \langle i'_m i'_{m+1} |
\end{equation}
as
\begin{equation}
\rho^{i_m i_{m+1}}_{i'_m i'_{m+1}} =c^\ast_{i_1 \ldots i'_m i'_{m+1} \ldots i_N} c_{i_1 \ldots i_m i_{m+1} \ldots i_N} \ , 
\end{equation}
where repeated indices  are implicitly summed over. Then, the constraint (\ref{eq:vanish_c}) imposes
\begin{equation}
i_m + i_{m+1} \neq i'_m + i'_{m+1} \longrightarrow \rho^{i_m i_{m+1}}_{i'_m i'_{m+1}} = 0 \ ,
\end{equation}
That is, we find
\begin{equation} 
 \rho^{00}_{01} = \rho^{00}_{10} = \rho^{00}_{11} = \rho^{01}_{11} = \rho^{10}_{11} = 0 \ ,
 \label{eq:den_zeros}
\end{equation}
which implies
\begin{equation}
 d=d'=f=e=e'=0.
\end{equation}
Note, that there is no restriction on the rest of elements of $\rho_{\left[m,m+1\right]}$. 

It is now necessary to impose that the ground state is not a generic eigenstate of the total $Z$-component of spin operator but, rather,
\begin{equation}
 \Lambda = 0 \ . \label{eq:eigen_zero}
\end{equation}
 However, no constraint on the 2-body reduced density matrix will follow from that unless 2-site translational invariance is assumed
\begin{equation}
 \rho_{\left[m,m+1\right]} = \rho_{\left[m+2,m+3\right]} \ \ \ \ \ \forall{m} \ .
 \label{eq:2_ti}
\end{equation}
Note that this condition is weaker than imposing complete translational symmetry, yet sufficient to constrain the elements of $\rho_{\left[m,m+1\right]}$
to obey
\begin{equation}
\tr \left( \left( \sigma_m^Z + \sigma_{m+1}^Z \right) \rho_{\left[m,m+1\right]} \right) = \Lambda = 0 \ \ \ \ \ \forall{m} \ .
\end{equation}
This further implies
\begin{equation}
\rho^{00}_{00} = \rho^{11}_{11} \equiv a \ . \label{eq:den_a}
\end{equation}

The whole set of constraints we have presented in 
 Eq. (\ref{eq:den_zeros}) and Eq. (\ref{eq:den_a}) reduce the number of parameters of the 2-body reduced density matrix (\ref{eq:twobody}) that, now, reads
\begin{equation}
\rho_{\left[m,m+1\right]} = \left( \begin{array} {cccc}
a & 0 & 0  & 0 \\
0 & b & c  & 0 \\
0 & c & b' & 0 \\
0 & 0 & 0  & a \\
\end{array} \right) \ .
\label{eq:twobody_eig}
\end{equation}
Note that this form corresponds to the density matrix for even sites, $m=2p$, whereas for odd sites, $m=2p-1$, coefficients $b$ and $b'$ should be exchanged. Furthermore, we can compute the 1-body reduced density matrix by tracing out one spin from the 2-body one
\begin{equation}
 \rho_{\left[m\right]} = \left( \begin{array} {cc}
   a +b & 0 \\
   0 & b'+a \\
   \end{array} \right) \ , 
\label{eq:onebody_eig}
\end{equation}
which is diagonal by construction.

%%%%%%%%%%%%%%%%%%%%%%%%%%%%%%%%%%%%
\subsubsection{Strict Translational Invariance}
The constraint of translational invariance 
produces a different form of the ansatz.
For instance,  
on the reduced density matrices, this condition implies
\begin{equation}
\rho_{\left[m,m+1\right]} = \rho_{\left[m+1,m+2\right]} \label{eq:1_ti} \longrightarrow
\rho^{i_mi_{m+1}}_{i_m' i_{m+1}'} = \rho^{i_{m+1}i_m}_{i_{m+1}' i_m'} \ ,
\end{equation}
that is,
\begin{equation}
\begin{split}
&\rho^{01}_{01} = \rho^{10}_{10} \equiv b \\
&\rho^{00}_{01} = \rho^{00}_{10} \equiv d \\
&\rho^{01}_{11} = \rho^{10}_{11} \equiv e \ . \\ 
\end{split}
\label{eq:den_ti}
\end{equation}
Consequently, the generic expression in Eq. (\ref{eq:twobody}) for $\rho_{\left[m,m+1\right]}$ will be casted into the form
\begin{equation}
\rho_{\left[m,m+1\right]} = \left( \begin{array} {cccc}
a & d & d & f \\
d & b & c & e \\
d & c & b & e \\
f & e & e & a' \\
\end{array} \right) \ \ \ \ \forall{m} \ . \label{eq:twobody_tr}
\end{equation}
Note that translational invariance reduces the set of parameters 
in a quite different way as compared to rotational invariance. 

The 1-body reduced density matrix can be obtained by integrating
out one of the spins in Eq. (\ref{eq:twobody_tr})
\begin{equation}
\rho_{\left[m\right]} = \left( \begin{array} {cc}
a+b & d+e \\
d+e & b+a' \\
\end{array} \right) \ . \label{eq:onebody_tr}
\end{equation}
This result remains the same whatever spin is chosen
to be integrated out, $\rho_{\left[m\right]} = \rho_{\left[m+1\right]}$.
Note that off-diagonal terms are now present, at variance with the
rotational invariant case. Therefore, these off-diagonal terms 
can be seen as a signature of the
lack of rotational symmetry in the state.

%%%%%%%%%%%%%%%%%%%%%%%%%%%%%%%%%%%%
\subsubsection{Quasi Translational Invariance}
The ground state of the Heisenberg spin chain turns out to verify
translational invariance up to a sign. That is, the ground state is invariant
under translating all spins by one position if, simultaneously, all of them
are flipped. This property is rooted at the combination of the anti-ferromagnetic
character of the Heisenberg interaction and the 2-body nearest neighbor
decomposition of the Hamiltonian. For the sake of simplicity, 
we call this symmetry "quasi translational
invariance". It is, then, necessary to modify our discussion on translational symmetry in order to accommodate this effect. It is easy to see
that the generic restriction stated by Eq. (\ref{eq:den_ti}) gets modified 
in the following way
\begin{equation}
\begin{split}
&\rho^{01}_{01} = \ \ \rho^{10}_{10} \equiv b \\
&\rho^{00}_{01} = -\rho^{00}_{10} \equiv d \\
&\rho^{01}_{11} = -\rho^{10}_{11} \equiv e \ . \\ 
\end{split}
\label{eq:den_ti_relax}
\end{equation}
To be precise, some off-diagonal terms include a minus sign. Let us note that the expectation value of the energy does not depend on this sign. Therefore, the new form of the 2-body reduced density matrix reads
\begin{equation}
 \rho_{\left[m,m+1\right]} = \left( \begin{array} {cccc}
 a & d & -d & f \\
 d & b & c & e \\
 -d & c & b & -e \\
 f & e & -e & a' \\
 \end{array} \right) \ , \label{eq:twobody_ti_relax}
\end{equation}
where $d \leftrightarrow -d$, $e \leftrightarrow -e$ should be exchanged for odd/even values of $m$. 
The 1-body reduced density matrix is also different for odd/even values of $m$
\begin{equation}
\rho_{\left[m\right]} = \left( \begin{array} {cc}
a+b & -d+e \\
-d+e & b+a' \\
\end{array} \right) \ , \label{eq:onebody_tr_relax}
\end{equation}
with a sign exchange for $d$ and $e$. Note that, again,
off-diagonal terms are present, indicating that
no rotational symmetry is preserved by this ansatz in general.

%%%%%%%%%%%%%%%%%%%%%%%%%%%%%%%%%%%%
\subsubsection{Exact ground state}
Finally, instead of considering symmetries separately, let us impose all rotational and quasi translational invariances simultaneously on a generic state. This is the relevant case of the exact ground state
of the Heisenberg model.
Then, the density matrix form is even simpler. From the combination of Eq. (\ref{eq:den_zeros}), Eq. (\ref{eq:den_a}) and Eq. (\ref{eq:den_ti_relax}) we  obtain the expression for the 2-body density matrix of the Heisenberg ground state {\cite{Wooters:2004}}
\begin{equation}
\rho_{\left[m,m+1\right]} = \left( \begin{array} {cccc}
a & 0 & 0 & 0 \\
0 & b & c & 0 \\
0 & c & b & 0 \\
0 & 0 & 0 & a \\
\end{array} \right) \ \ \ \ \forall{m} \ . \label{eq:twobody_hh}
\end{equation}
Only a few parameters $\lbrace a, b, c \rbrace$ remain undetermined by
symmetries.

The 1-body reduced density matrix turns out to be
\begin{equation}
\rho_{\left[m\right]} = \left( \begin{array} {cc}
1/2 & 0 \\
0 & 1/2 \\
\end{array} \right) \ \ \ \ \ \ \forall{m} \ 
\label{eq:onebody_hh}
\end{equation}
as a consequence of the unitarity of this reduced density matrix.
This reflects the fact that each individual spin in the system has equal probability to
take any value in any direction since $\tr \left( \sigma_Z \rho_m \right) =0$ ($Z\leftrightarrow X$,
$Z\leftrightarrow Y$), as consequence of rotational {\sl and} translational
invariance. Note that there is no distinction here between the constraints
coming from strict translational invariance or quasi translational invariance. 

Later on, we shall use the 1-body and (adjacent) 2-body reduced density matrices as figures of merit to determine
how rotational symmetry or quasi translational invariance are maintained within the MPS approximation.

%%%%%%%%%%%%%%%%%%%%%%%%%%%%%%%%%%%%%%%%%%%%%%%%%%%%%%%%%%%%%%%%%%%%%%%%%%%%%%
%%%%%%%%%%%%%%%%%%%%%%%%%%%%%%%%%%%%%%%%%%%%%%%%%%%%%%%%%%%%%%%%%%%%%%%%%%%%%%
%%% Resultat NUMERIC amb MPS
\section{Matrix Product States}
\label{Sect:iTEBD}

\begin{figure}[h]
  \resizebox{6cm}{!}{\includegraphics{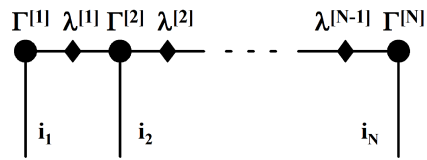}}
  \caption{MPS diagrammatic representation. Each filled circle represents a $\Gamma$ and each line represents an index. The diamonds represent the Schmidt coefficients $\lambda$. }
\label{fig:MPS}
\end{figure}

%MPS representation.
The Matrix Product State ansatz is a powerful tool to describe one-dimensional systems with a reduced number of parameters. The basic idea underlying this approximation is to
represent the set of $2^N$ coefficients that describe any state of  a $N$-spin chain as a product of  $N$ matrices that can be chosen as follows \cite{Vidal:MPS}
\begin{equation}
  | \psi \rangle = \sum_{i_1 \ldots i_N = 1}^2 \Gamma^{[1]i_1} \lambda^{[1]} \ldots \lambda^{[N-1]} \Gamma^{[N]i_N} |i_1
  \rangle \ldots |i_N \rangle \ .
\label{Eq:state}
\end{equation}
All $\Gamma^{[m]i_m}$ objects are $\chi \times \chi$ matrices and $\lambda^{[m]}$ are diagonal matrices of dimension $\chi$. The state can be pictured
as shown in Fig. \ref{fig:MPS}.

The rational for this ansatz relies on the fact that the above construction
is the result of a systematic use of Schmidt decompositions. That is, at any
site $k$ we may divide the spin chain in two pieces. The state can then be
written as
\begin{equation}
  \label{Eq:schmidt}
  \vert \psi\rangle=\sum_{\alpha=1}^{\chi}\vert Left_\alpha\rangle \vert Right_\alpha\rangle \ ,
\end{equation}
where $\vert Left_\alpha\rangle$ and $\vert Right_\alpha\rangle$ stand for the states that form an orthogonal basis on the left
and right parts of the chain. 

As a consequence of this Schmidt decomposition, several measures of entanglement can be constructed. The most elementary one corresponds to
the value of $\chi$ itself, called Schmidt number. 
For instance, $\chi=1$ on a given link means that the state can be
divided there as a simple product of left and right states. In such a case, there is no
entanglement between the left and right parts of the system. 
The opposite case of maximum   $\chi= 2^\frac{N}{2}$ corresponds to
maximally entangled states. From the Schmidt decomposition we can
also construct the Von Neumann entanglement entropy
\begin{equation}
S = -\sum_{\alpha=1}^{\chi}{\lambda_{\alpha}^2\log \lambda_{\alpha}^2 } \ .
\end{equation}
Zero entropy corresponds to uncorrelated partitions, whereas a linear scaling
of entropy with the number of sites in the smaller partition corresponds to
maximal entanglement.

Extensive efforts have been devoted to understand the amount of entanglement
which is present in the ground states of strongly correlated systems. 
At quantum phase transitions, the entanglement entropy for a block of adjacent spins scales logarithmically with the number of sites for critical models
\cite{VLRK}. 
The ground state of the Heisenberg spin chain displays this scaling property.

Note that the Schmidt decomposition produces naturally the following isometries
\begin{equation}
\sum_{i,\alpha}\lambda_{\alpha}^{[m-1]} \Gamma_{\alpha \beta}^{ \left[ m \right] i} \left( \lambda_{\alpha}^{[m-1]} \Gamma_{\alpha \beta'}^{\left[ m \right]i} \right)^* = \delta_{\beta \beta'} \label{eq:isom_left}
\end{equation}
and
\begin{equation}
\sum_{i,\beta} \Gamma_{\alpha \beta}^{ \left[ m \right] i} \lambda_{\beta}^{[m]}  \left( \Gamma_{\alpha' \beta}^{\left[ m \right]i} \lambda_{\beta}^{[m]} \right)^* = \delta_{\alpha \alpha'} \label{eq:isom_right} \ .
\end{equation}
This is an instrumental property to simplify the contractions
of states that have a MPS structure.

This approach is closely related to the DMRG ansatz. Note that the DMRG algorithm produces
a MPS state, that is,
a state whose coefficients are represented as a product of matrices. 
This can be translated into the form in Eq. (\ref{Eq:state})
using a local symmetry. Indeed, at any point between two sites
in the product of matrices,
we can insert $I=XX^{-1}$ and reabsorb $X$ and $X^{-1}$
on the adjacent left and right matrices respectively. This
(gauge) freedom can be used to convert 
the form of the state in such a way that Eq. 
(\ref{eq:isom_left}) and Eq. (\ref{eq:isom_right}) are fulfilled.

The challenge of finding the best set of matrices $\left \lbrace \Gamma^{[m]i_m} \right \rbrace$, $\left \lbrace \lambda^{[m]} \right \rbrace$ has already been faced using numerical algorithms. We shall here use the iTEBD algorithm proposed
in Ref. \cite{Vidal:iTEBD}. The basic idea consists in evolving some initial state in imaginary time.  Provided that its overlap with the ground state is different from zero, the imaginary time evolution converges towards
the ground state of the system. This strategy can be implemented
using the Trotter decomposition \cite{Trotter} that allows to decompose
the imaginary time evolution of the total Hamiltonian in small
steps, each one corresponding to the imaginary time
evolution of a piece of the Hamiltonian.
Moreover, whenever the Hamiltonian is made of two-body nearest-neighbor interactions, it can be written as a sum of two non-commuting pieces which are made of commuting terms. Within each non-commuting subspace, the action of the Hamiltonian on the evolution of the state reads as a product of exponentials acting on two adjacent sites. Consequently, the state can be described by only two different matrices \cite{Vidal:iTEBD}
\begin{equation}
\begin{split}
&\Gamma^{[2m-1]} \equiv \Gamma^{[A]} \ , \ \ \Gamma^{[2m]}  \equiv \Gamma^{[B]}\quad \\
&\lambda^{[2m-1]}  \equiv \lambda^{[A]} \ , \ \ \lambda^{[2m]}  \equiv \lambda^{[B]}\quad \\
\end{split}
\ \ \ \forall m \ .
\end{equation}
It follows that the full algorithm reduces to a two-step iteration over two sites. 

Let us emphasize that the iTEBD algorithm distinguishes as
many sites as the number of terms involved in the local interaction.
For instance, a Hamiltonian made with interactions that involve
next-to-nearest-neighbor sites would be handled by iTEBD using
three type of matrices $A$, $B$ and $C$ and a three-step iteration euclidean evolution.
Note as well that, in our case, the iTEBD algorithm incorporates explicitly two-site translational invariance.

The choice for the parameter $\chi$ of the MPS, that is the
size of the matrices in the ansatz (often called $m$ in DMRG literature) 
controls both the computational cost and accuracy of the approximation.
Some results for the energy and the entropy of the ground state
of the Heisenberg spin-1/2 infinite chain obtained with the iTEBD algorithm 
as a function of $\chi$ are  
shown in Table \ref{tab:t1}.

%\begin{table}
\begin{table}[h]
\centering
\begin{tabular}[b]{|c|c|c|}
\hline
$\chi$ & $E_{\chi}$ & $S_{\chi}$ \\
\hline
2 & -0.427908 & 0.48570 \\
3 & -0.435784 & 0.60866 \\
4 & -0.441058 & 0.76413 \\
5 & -0.442048 & 0.84778 \\
\hline
$\infty$ & -0.44314718 & $\infty$ \\
\hline
\end{tabular}
\caption[]{Comparison between the exact result (labeled as
$\chi=\infty$) and the $\chi$-dependent iTEBD/DMRG approximation for the energy $E_{\chi}$ and the Von Neumann entropy $S_{\chi}$ in the Heisenberg spin-1/2 infinite chain. All numbers in the table are precise up to the last figure.} \label{tab:t1}
\end{table}

%%%%%%%%%%%%%%%%%%%%%%%%%%%%%%%%%%%%%%%%%%%%%%%%%%%%%%%%%%%%%%%%%%%%%%%%%%%%%%
%%%%%%%%%%%%%%%%%%%%%%%%%%%%%%%%%%%%%%%%%%%%%%%%%%%%%%%%%%%%%%%%%%%%%%%%%%%%%%
%%% Resultat EXACTE amb l'aproximacio MPS
\section{Exact Results for MPS}
\label{Sect:exact}
In this section we shall present an exact method for finding the 
MPS approximation to the ground state of the Heisenberg Hamiltonian. Then, we shall study the lowest-$\chi$ cases in detail.

%%%%%%%%%%%%%%%%%%%%%%%%%%%%%%%%%%%%
\subsection{Exact method}

%Degrees of freedom of Gamma's & lambda's.
Instead of resorting to a numerical algorithm in order to find the
MPS approximation to the ground state of the Heisenberg Hamiltonian, we shall construct the exact form of this approximation by exploiting the isometry constraints that the matrices $\left\lbrace \Gamma^{ \left [m \right ]i_m}, \lambda^{[m]} \right\rbrace$ obey. It is, thus, necessary to understand
the MPS approach as a sequence of Schmidt decompositions rather than
the result of a DMRG procedure. This is the key point to our
analytical strategy.

%Constraints
We first observe that the 2-body reduced density matrix for a MPS 
can be written in a simple form as a function of the matrices $\left \lbrace \Gamma^{\left[ A, B \right]i}, \lambda^{[A, B]} \right \rbrace$. That is,
the combined exploitation of two-site translational symmetry
and isometries implies
\begin{equation}
\begin{gathered}
\rho^{ij}_{i'j'} =  \\
\textrm{tr} \left( \lambda^{[B]} \Gamma^{[A]i} \lambda^{[A]} \Gamma^{[B]j} \lambda^{[B]} \lambda^{[B]} \Gamma^{\dagger[B]j'} \lambda^{[A]} \Gamma^{\dagger[A]i'} \lambda^{[B]} \right) \ . 
\label{eq:rho}
\end{gathered}
\end{equation}
Isometries are, indeed, necessary to seamlessly contract the 
infinite chain except for the open physical indices of the reduced density matrix.
In other words, the above expression is only valid in the gauge
where the matrices forming the MPS are appropriately chosen to
fulfill the isometry constraints in Eqs. (\ref{eq:isom_left},\ref{eq:isom_right}).

\begin{figure}[h]
  \resizebox{5cm}{!}{\includegraphics{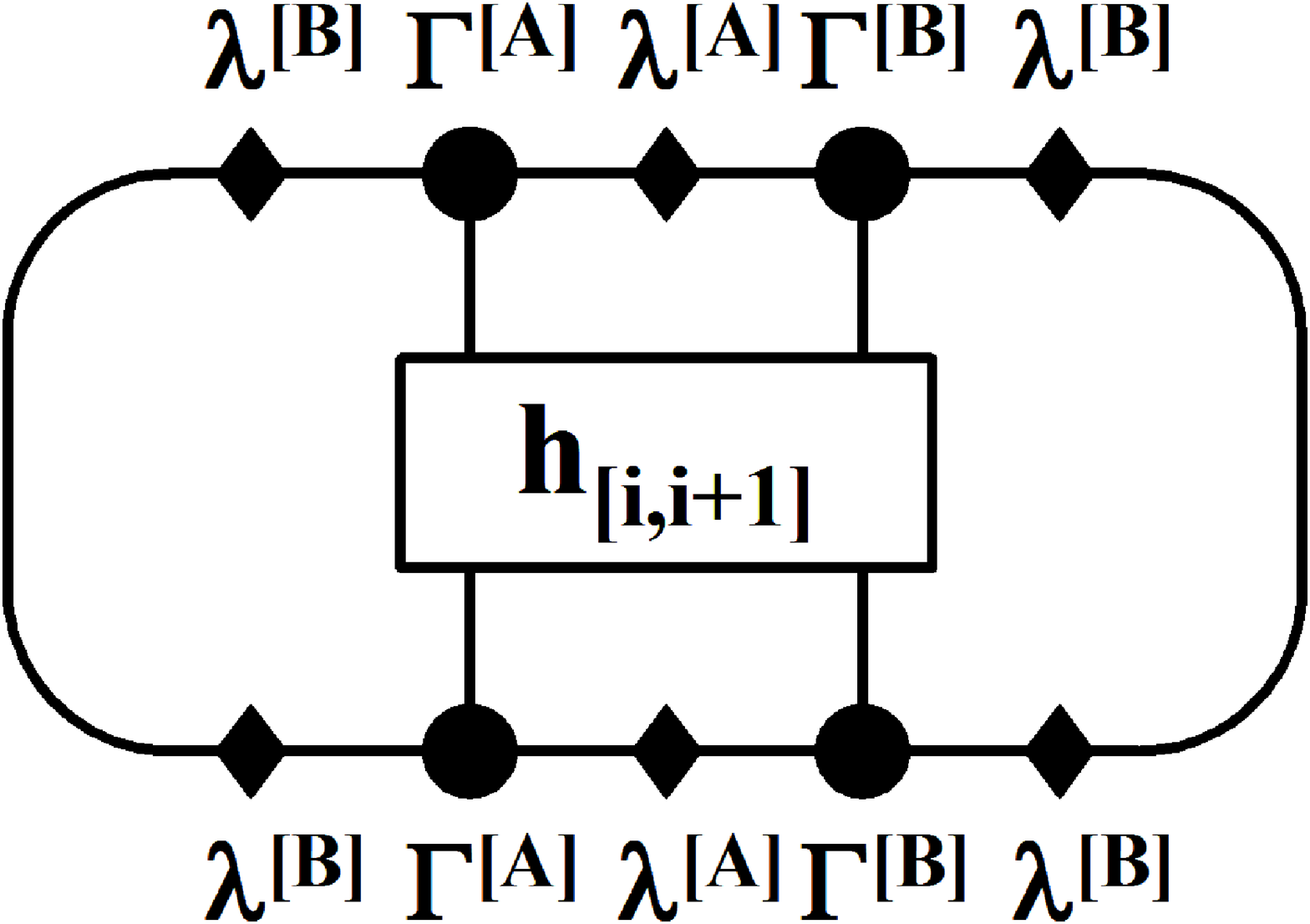}}
  \caption{Diagrammatic representation for the expected value of a local term of the Hamiltonian. Closed lines represent summed indices.} \label{fig:energy}
\end{figure}

%Energy
Once we know the dependence of the 2-body reduced density matrix on the MPS matrices, we can easily compute any observable. In particular, let us write down the energy of the system which is the quantity we want to minimize. 
First, we take a local term at sites $[m,m+1]$ for two spins in the Heisenberg Hamiltonian (\ref{eq:ham})
\begin{equation}
h_{\left[ m,m+1 \right]} = \frac{1}{4}\left ( \begin{array} {cccc}
1 & 0 & 0 & 0 \\
0 & -1 & 2 & 0 \\
0 & 2 & -1 & 0 \\
0 & 0 & 0 & 1 \\
\end{array} \right)
\end{equation}
and write its expected value using the form of the
2-body reduced density matrix in Eq. (\ref{eq:rho})
\begin{equation}
\left\langle h \right\rangle = {\rm tr}(\rho h)=\frac{1}{4} - \frac{1}{2} \textrm{tr} \left( \lambda^B M \left(\lambda^B M \right)^{\dagger} \right) \ ,
\end{equation}
where
\begin{equation}
M \equiv \Gamma^{[A]1} \lambda^{[A]} \Gamma^{[B]2} \lambda^{[B]} - \Gamma^{[A]2} \lambda^{[A]} \Gamma^{[B]1} \lambda^{[B]} \ .
\end{equation}
This expected value can also be written in a diagrammatic form, see Fig. \ref{fig:energy}.

At this point the set of isometries $\left\lbrace \Gamma^{[A]i}, \Gamma^{[B]i} \right\rbrace$ are still general $\chi \times \chi$ matrices satisfying Eq. (\ref{eq:isom_left}) and 
(\ref{eq:isom_right}). From now on we will impose some additional requirements on them so as to incorporate the symmetries of the problem. 

%More constraints...
For instance, it is possible to impose
\begin{equation}
\Gamma^{[A]i} = \pm \Gamma^{[B]i} \equiv \Gamma^i \ , 
\ \ \lambda^{[A]} = \lambda^{[B]} \equiv \lambda 
\label{eq:t_inv}
\end{equation}
in order to implement (quasi) translationally invariance. 
In such a case, it follows that
\begin{equation}
M_{\pm} = \Gamma^1 \lambda \Gamma^2 \lambda \mp \Gamma^2 \lambda \Gamma^1 \lambda \ .
\label{eq:M_ti}
\end{equation} 
We, then, conclude that the option of quasi translational invariance produces a lower energy than the strict translational invariant case. This, as mentioned
before, is related to the anti-ferromagnetic character of the Heisenberg
interaction.

Alternatively, we can explicitly break translational invariance and set
\begin{equation}
\Gamma^{[A]i} = \left( \pm \right)^j \Gamma^{[B]j} \equiv \Gamma^i , \ \ \ 
\lambda^{[A]} = \lambda^{[B]} \equiv \lambda\ \, \label{eq:non_t_inv}
\end{equation} 
where $i,j=\left\lbrace 1, 2 \right\rbrace$, $i \neq j$. Then, 
\begin{equation}
M_{\pm} = \pm \left( \Gamma^1 \lambda \Gamma^1 \lambda \mp \Gamma^2 \lambda \Gamma^2 \lambda \right) \ . 
\label{eq:M_nti}
\end{equation}
This case will later be seen on explicit $\chi$ examples to correspond to a necessary but not
sufficient condition to maintain rotational invariance. Again, in this case,
the minus sign is favored by minimization of the energy.

Both scenarios lead to the minimization of an expression 
of the form
\begin{equation}
\left\langle h \right\rangle = \frac{1}{4} - \frac{1}{2} \textrm{tr} \left( \lambda M_{-} \left( \lambda M_{-} \right)^{\dagger} \right) \ ,
\label{eq:maxim}
\end{equation}
where $M_-$ is given by either Eq. (\ref{eq:M_ti}) or Eq. (\ref{eq:M_nti}).
In order to make further analytical progress, we now turn to 
the particular cases of low values for $\chi$.

%%%%%%%%%%%%%%%%%%%%%%%%%%%%%%%%%%%%
\subsection{\boldmath{} $\chi=2$ \unboldmath{}}
Let us now focus on the simplest case of MPS with $\chi=2$. We can exploit 
in different ways the freedom in implementing the isometry conditions on matrices $\Gamma$ and $\lambda$ to simplify further the MPS ansatz. We shall
discuss in turn several possibilites that deliver the same
minimum energy, though not the same state.

\subsubsection{Eigenstate of the total $Z$-component of spin}
As discussed previously, the constraint in Eq. (\ref{eq:non_t_inv}) allows for the choice of (up to a sign) identical matrices. The isometry property
leaves enough space to further restrict the form of the these matrices without loss of generality
to be
\begin{gather}
\Gamma^{\left[ A \right]1} = \left( \begin{array} {cc} a & 0 \\ 0 & 0 \\ \end{array} \right) \\ \Gamma^{\left[ A \right]2} = \left( \begin{array} {cc} 0 & b \\ b & 0 \\ \end{array} \right) \\ \label{eq:gamma2_nti}
\lambda^A = \lambda^B = \left( \begin{array} {cc} \lambda_1 & 0 \\ 0 & \lambda_2 \end{array} \right) \ .
\end{gather}
Then, the expression for the energy takes the form
\begin{equation}
\left\langle h \right\rangle = \frac{1}{4} - \frac{1}{2} \left( b^4 \lambda_1^2 \lambda_2^4 + \lambda_1^4 \left( a^2 \lambda_1 + b^2 \lambda_2 \right)^2 \right) ,
\end{equation}
and the isometry condition reduces to
\begin{equation}
b^2 \lambda_1^2 = 1 \ \ , \ \ a^2 \lambda_1^2 + b^2 \lambda_2^2 = 1 \ .
\end{equation}
These isometry conditions can be solved explicitly
\begin{equation}
b^2=\frac{1}{\lambda_1^2}\ , \ a^2=\frac{2 \lambda_1^2-1}{\lambda_1^4}\ ,
\end{equation}
where we have used the normalization condition $\lambda_1^2 + \lambda_2^2 = 1 $.
The energy of the system can, then, be written as a unique function of $\lambda_1$ 
\begin{equation}
\left\langle h \right\rangle = \frac{11}{4} - 2 \lambda_1^2 - \frac{1}{\lambda_1^2} - 2 \lambda_1 \sqrt{1-\lambda_1^2}+ \frac{\sqrt{1-\lambda_1^2}}{\lambda_1} \ . \label{eq:energy_l}
\end{equation}
After some algebraic manipulations are performed, 
the minimum of this expression can be reduced  to
\begin{equation}
4 + 3\lambda_1^2-14 \lambda_1^4-8 \lambda_1^6+16\lambda_1^8=0 \ .
\end{equation}
One of the solutions of this quartic equation produces the lowest energy at
\begin{equation}
\lambda_{1,min} = 0.94586362...
\end{equation}
\begin{equation}
 E_{\chi=2} = -0.42790801...
\end{equation}
which is the same minimum obtained with numerical techniques (see
Table \ref{tab:t1}). This result confirms that DMRG and iTEBD
find numerically the correct minimum for the energy which is
consistent with a MPS ansatz.

%Transfer Matrix & Correlation Length
We can also calculate the correlation length of the MPS from the transfer matrix $E = \sum_i \left( \Gamma^i \lambda \right) \otimes \left( \Gamma^{i\ast} \lambda^{\ast} \right)$. In this particular case, $E$ can be written as a function of $\lambda_1$:
\begin{equation}
E = \left( \begin{array} {cccc}
0 & 0 & 0 & 1 \\
0 & 0 & \frac{\sqrt{1-\lambda_1^2}}{\lambda_1} & 0 \\
0 & \frac{\sqrt{1-\lambda_1^2}}{\lambda_1} & 0 & 0 \\
\frac{1}{\lambda_1^2} - 1 & 0 & 0 & 2 - \frac{1}{\lambda_1^2} \\
\end{array} \right)
\end{equation}
whose eigenvalues define the correlation length $\xi = 1/\log \left( \frac{e_1}{e_2} \right)$. The explicit dependence of $\xi$ on $\lambda_1$ is
\begin{equation}
\xi^{-1}_{\chi=2} \left( \lambda_1 \right) = \log \frac{2\lambda_1^2 -1 + \sqrt{8\lambda_1^2 \left( \lambda_1^2-1 \right) +1}}{2\lambda_1 \sqrt{1-\lambda_1^2}} \ .
\end{equation}
Setting $\lambda_1 = \lambda_{1,min}$ we find
\begin{equation}
\xi_{\chi=2} = 0.93820167...
\end{equation}
The exact value of the half chain entropy for $\chi=2$ turns out to be
\begin{equation}
S_{\chi=2}=0.48570420...
\end{equation}

Finally, following the notation in Eq. (\ref{eq:twobody_eig}) we can also write down the 2-body density matrix of this MPS
\begin{equation}
\begin{split}
&a = 3-\frac{1}{\lambda_1^2}-2\lambda_1^2 \\
&b = -4+\frac{1}{\lambda_1^2}+4\lambda_1^2 \\
&b'= -1+\frac{1}{\lambda_1^2} \\
&c = \left( \frac{1}{\lambda_1} -2\lambda_1 \right) \sqrt{1-\lambda_1^2} \\
\end{split}
\end{equation}
and set $\lambda_1 = \lambda_{1,min}$
\begin{equation}
\rho_{\left[ m,m+1 \right]} = \left( \begin{array} {cccc}
0.0929 & 0 & 0 & 0 \\
0 & 0.6964 & -0.2708 & 0 \\
0 & -0.2708 & 0.1177 & 0 \\
0 & 0 & 0 & 0.0929 \\
\end{array} \right) \ .
\end{equation}
Tracing out one spin, the 1-body density matrix becomes
\begin{equation}
\rho_{\left[ m \right]} = \left( \begin{array} {cc}
-1+2\lambda_1^2 & 0 \\
0 & 2 - 2\lambda_1^2 \\
\end{array} \right) \ ,
\end{equation}
that yields
\begin{equation}
\rho_{\left[m\right]} = \left( \begin{array} {cc}
0.7893 & 0 \\
0 & 0.2107 \\
\end{array} \right) \ .
\end{equation}

Note that the chosen parametrization leads to an eigenstate of the total $Z$-component of spin (because of the block-diagonal form of the density matrices). However, the translational invariance is explicitly broken since $\rho_{\left[m-1,m\right]} \neq \rho_{\left[m,m+1\right]}$ and also $\rho_{\left[m\right]} \neq \rho_{\left[m+1\right]}$.

\subsubsection{Quasi Translational Invariance}
A second possibility we can follow corresponds to setting conditions (\ref{eq:t_inv}). In that case, we try to find out the minimum energy state with matrices $\Gamma$ parametrized as
\begin{equation}
\Gamma^{\left[ A \right]1} = \left( \begin{array} {cc} a & b \\ -b & 0 \\ \end{array} \right) \ \ \Gamma^{\left[ A \right]2} = \left( \begin{array} {cc} a & -b \\ b & 0 \\ \end{array} \right) \ ,
\end{equation}
and equal $\lambda$ matrices for all sites. 
This new parametrization can be seen to produce the very same dependence of the energy on $\lambda_1$. Therefore,  the minimum energy we shall
obtain is the same than the first parametrization reached. Furthermore, the computation of the correlation length leads us to the same result as before. Despite these coincidences, the state is no more the same, which can be inferred from the 2-body density matrix of the system:
\begin{equation}
\begin{split}
&a = \frac{1}{4}+\frac{\sqrt{1-\lambda_1^2}}{2\lambda_1}-\lambda_1 \sqrt{1-\lambda_1^2} \\
&b = \frac{1}{4}-\frac{\sqrt{1-\lambda_1^2}}{2\lambda_1}+\lambda_1 \sqrt{1-\lambda_1^2} \\
&c = \frac{11}{4}-\frac{1}{\lambda_1^2}-2 \lambda_1^2+\frac{\sqrt{1-\lambda_1^2}}{2 \lambda_1}-\lambda_1 \sqrt{1-\lambda_1^2} \\
&d = e' = \frac{3}{4}-\lambda_1^2 \\
&d' = e = -d = -e' \\
&f = \frac{11}{4}-\frac{1}{\lambda_1^2}-2 \lambda_1^2-\frac{\sqrt{1-\lambda_1^2}}{2\lambda_1}+\lambda_1 \sqrt{1-\lambda_1^2} \ . \\
\end{split}
\end{equation}
Note that the equalities $d=d'$, $e=e'$ are maintained up to a sign, since the condition we require is precisely quasi translational invariance. Numerically, the 2-body reduced density matrix and also the 1-body reduced density matrix take the form:
\begin{equation}
\rho_{\left[m,m+1\right]} = \left( \begin{array} {cccc}
 0.1146 & -0.1447 &  0.1447 & -0.0216 \\
-0.1447 &  0.3854 & -0.2925 &  0.1447 \\
 0.1447 & -0.2925 &  0.3854 & -0.1447 \\
-0.0216 &  0.1447 & -0.1447 &  0.1146 \\
\end{array} \right)
\end{equation}
\begin{equation}
\rho_{\left[m\right]} = \left( \begin{array} {cc}
0.5 & 0.2893 \\
0.2893 & 0.5 \\
\end{array} \right) \ .
\end{equation}

The state represented by this MPS is not an eigenstate of the total $Z$-component of spin, but quasi translational invariance is recovered.
The two parametrizations we have analyzed deliver the same energy but
handle differently the original symmetries of the problem. The resulting
MPS will give different results on local observables attached to these symmetries. We may also consider a MPS which is not forced to take
the form imposed by any symmetry. Such new state would still deliver
the same energy and differ on other observables.

%%%%%%%%%%%%%%%%%%%%%%%%%%%%%%%%%%%%
\subsection{\boldmath{} $\chi \geq 3$ \unboldmath{}} 
We can also follow similar steps for higher values of $\chi$. In particular, let us apply the constraint discussed in Eq. (\ref{eq:non_t_inv}) for matrices of $\chi=3$. This requirement allows us to choose a rather simple form of the matrices
\begin{equation}
\Gamma^{\left[ A \right]1} = \left( \begin{array} {ccc} a & 0 & c \\ 0 & 0 & 0 \\ c & 0 & e \\ \end{array} \right) \ \ \Gamma^{\left[ A \right]2} = \left( \begin{array} {ccc} 0 & b & 0 \\ b & 0 & d \\ 0 & d & 0 \\ \end{array} \right) \ 
\end{equation}
which generalizes the form of matrices $\Gamma$ in Eq. (\ref{eq:gamma2_nti}) for $\chi=2$.

Note that the number of parameters that describes the state can be reduced by considering the isometry and normalization conditions
\begin{equation}
\begin{split}
&a^2\lambda_1^2 + b^2\lambda_2^2 + c^2\lambda_3^2 = 1 \\
&b^2\lambda_1^2 + d^2\lambda_3^2 = 1 \\
&^2\lambda_1^2 + d^2\lambda_2^2 + e^2\lambda_3^2 = 1 \\
&ac\lambda_1^2 + bd\lambda_2^2 + ce\lambda_3^2 = 0 \ . \\
\end{split}
\end{equation}
As a consequence of this reduction, the energy of the state can be expressed, after some algebra, as a function of only three parameters. Therefore, the problem of finding a good approximation to the ground state of the Heisenberg Hamiltonian turns into a three-variable function minimization. The solution is again the same as numerical algorithms reach with no constraint on the matrices
\begin{equation}
E_{\chi=3} = -0.435784... \label{eq:sol_chi3}
\end{equation}
and the reduced density matrices take the form
\begin{gather}
\rho_{\left[m,m+1\right]} = \left( \begin{array} {cccc}
0.0930 & 0 & 0 & 0 \\
0 & 0.6429 & -0.2788 & 0 \\
0 & -0.2788 & 0.1711 & 0 \\
0 & 0 & 0 & 0.0930 \\
\end{array} \right) \\
\rho_{\left[m\right]} = \left( \begin{array} {cc}
0.7359 & 0 \\
0 & 0.2641 \\
\end{array} \right) \ .
\end{gather}

%Translational Invariance is violated.
We observe that again the $2(1)$-body reduced density matrix of the MPS clearly violates translational invariance. Nevertheless, in analogy to the $\chi=2$ case, we can require quasi translational invariance conditions on matrices $\Gamma$, such that Eq. (\ref{eq:t_inv}) is fulfilled. In this case, we have found the particular form
\begin{equation}
\Gamma^{\left[ A \right]1} = \left( \begin{array} {ccc} a & b & c \\ -b & 0 & d \\ c & -d & e \\ \end{array} \right) \ \ \Gamma^{\left[ A \right]2} = \left( \begin{array} {ccc} a & -b & c \\ b & 0 & -d \\ c & d & e \\ \end{array} \right)
\end{equation}
which produces the solution (\ref{eq:sol_chi3}) as well. However, despite obtaining the same minimum, the MPS is different from the previous one. This can be explicitly seen from the density matrices
\begin{gather}
\rho_{\left[m,m+1\right]} = \left( \begin{array} {cccc}
 0.1106 & -0.1180 &  0.1180 & -0.0176 \\
-0.1180 &  0.3894 & -0.2964 &  0.1180 \\
 0.1180 & -0.2964 &  0.3894 & -0.1180 \\
-0.0176 &  0.1180 & -0.1180 &  0.1106 \\
\end{array} \right) \\
\rho_{\left[m,m+1\right]} = \left( \begin{array} {cc}
0.5 & 0.2359 \\
0.2359 & 0.5 \\
\end{array} \right) \ .
\end{gather}
In this second case, quasi translational symmetry is preserved, but the rotational invariance is clearly broken. 

A similar reasoning can be followed in regard to higher $\chi$ MPS, looking for a generalized form of matrices $\Gamma$ and addressing the minimization of the energy by reducing the number of parameters that describes the MPS.
We shall quote some of the results obtained in the case of higher $\chi$ in the next section.

%%%%%%%%%%%%%%%%%%%%%%%%%%%%%%%%%%%%%%%%%%%%%%%%%%%%%%%%%%%%%%%%%%%%%%%%%%%%%%
%%%%%%%%%%%%%%%%%%%%%%%%%%%%%%%%%%%%%%%%%%%%%%%%%%%%%%%%%%%%%%%%%%%%%%%%%%%%%%
\section{Symmetry Restoration}
\label{Sect:symmetry}
The faithfulness of the MPS approximation to the ground state of a given Hamiltonian depends on the value of parameter $\chi$. We refer the reader to the extensive analysis of scaling relations 
performed in Refs. \cite{TOIL:2008,andersson}. It is, there, 
shown that
the finiteness of $\chi$  produces an effective
correlation length in the system
\begin{equation}
  \xi=\chi^\kappa \label{eq:scaling}
\end{equation} 
with $\kappa=1.37(1)$ for the Heisenberg spin chain. In the present case, we shall 
analyze a different effect, namely the breaking of the symmetries of the problem
as a function of $\chi$.

As discussed along the previous section, the MPS ansatz does not preserve all the symmetries of the exact state to be represented. Therefore, the symmetry breaking generated by this approximation should also be described as a function of $\chi$. In particular, for the limit $\chi \rightarrow \infty$ all the symmetries of the state described by the MPS should be completely restored. 

In order to quantify the symmetry breaking of a MPS with parameter $\chi$, we need to construct a figure of merit from the representation of the state. 
We shall present such a construction via an example. Let  us write down a
generic  MPS for $\chi=2$ obtained with the iTEBD algorithm with no restriction on the form of $\Gamma$ matrices
\begin{gather}
\rho_{\left[m,m+1\right]} = \left( \begin{array} {cccc}
 0.0972 & -0.0725 &  0.0554 & -0.0042 \\
-0.0725 &  0.6623 & -0.2751 &  0.0725 \\
 0.0554 & -0.2751 &  0.1433 & -0.0554 \\
-0.0042 &  0.0725 & -0.0554 &  0.0972 \\
\end{array} \right) \\
\rho_{\left[m\right]} = \left( \begin{array} {cc}
0.7595 & 0.1279 \\
0.1279 & 0.2405 \\
\end{array} \right) \ .
\end{gather}
Note that these reduced density matrices cast into the form stated by Eqs. (\ref{eq:twobody},\ref{eq:onebody}). In particular, let us focus on the 1-body reduced density matrix and write it as
\begin{equation}
\rho_{\left[m\right]} = \left( \begin{array} {cc}
\frac{1}{2}+x & y \\
y & \frac{1}{2}-x \\
\end{array} \right) \label{eq:1body_par}
\end{equation}
where the property $\tr \left( \rho_{\left[m\right]} \right) = 1$ has been used.

We see that the 1-body reduced density matrix is described by only two parameters $\left \lbrace x, y \right \rbrace$ that parametrize the deviation from the exact ground state. Indeed, with this notation, the 1-body reduced density matrix of the ground state of the Heisenberg chain corresponds to the setting $\left \lbrace x=0, y=0 \right \rbrace$ (see Eq. (\ref{eq:onebody_hh})). Therefore, these parameters account for the symmetries breaking of the MPS approximation.

This short description allows to compare again the two regimes studied above, namely the preservation of either rotational symmetry or quasi translational invariance. Let us summarize in Table \ref{tab:t2} the results of the previous section together with the generic 1-body reduced density matrix obtained with the iTEBD algorithm for $\chi=2$.

\begin{table}[h]
\centering
\begin{tabular}[c]{|c|c|c|c|}
\hline
\ & \textit{Gen} & \textit{Rot} & \textit{qTr} \\
\hline
$x$ & 0.2595 & 0.2893 & 0 \\
$y$ & 0.1279 & 0 & 0.2893 \\
$\Delta\mu\equiv 2\sqrt{x^2+y^2}$&0.2893&0.2893&0.2893\\
\hline
\end{tabular}
\caption[]{Comparison between the parameters of the 1-body reduced density matrix (see Eq. (\ref{eq:1body_par})) of a MPS with $\chi=2$ in the following cases: generic (\textit{Gen}), rotational invariance (\textit{Rot}) and quasi translational invariance (\textit{qTr}). Note that $\Delta\mu$ is identical
in all cases as it corresponds to the unique invariant coming from the
eigenvalues of the 1-body reduced density matrix.} 
\label{tab:t2}
\end{table}

The results from Table \ref{tab:t2} show explicitly the differences between the states represented by MPS with different symmetries preserved. Nevertheless, all of them correspond to the very same values for the energy and the entropy. In such a case, we should be able to construct from the 1-body reduced density matrix an observable that shall not depend on the gauge.

Since the elements of the density matrix depend explicitly on the local basis, the unique parameters that are invariant under local unitary transformations shall be its eigenvalues, that we call $\left \lbrace \mu_1,\mu_2 \right \rbrace$. Furthermore, note that the trace of the 1-body reduced density matrix is also constrained to be equal to one. Therefore, the only gauge-invariant parameter that remains is the difference of the eigenvalues
\begin{equation}
\Delta \mu \equiv \mu_1 - \mu_2 = 2\sqrt{x^2+y^2}
\end{equation}
which can be read explicitly as a function of parameters $\left \lbrace x,y \right \rbrace$.

\begin{figure}[h]
  \resizebox{8cm}{!}{\includegraphics{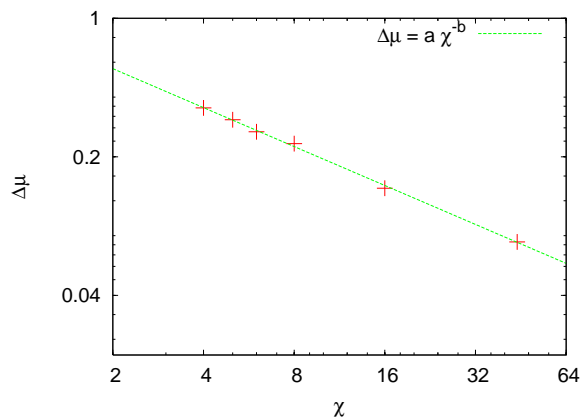}}
  \caption{Restoration of the broken symmetries parametrized by 
the difference of eigenvalues $\Delta\mu=\mu_1-\mu_2$ of the 1-body reduced density matrix gauge-invariant parameter.}
\label{fig:delta_mu}
\end{figure}

The quantity $\Delta \mu$ does not depend on the gauge and accounts for the faithfulness of the approximation. The freedom in implementing the isometry conditions on the matrices of the MPS allows to choose a particular form in which this approximation is performed. This choice is eventually a particular parametrization $\left \lbrace x,y \right \rbrace$ of the same amount of symmetry breaking $\Delta \mu$. Furthermore, this invariant observable shall depend on the parameter of the approximation, $\chi$. Some results are shown in Fig. \ref{fig:delta_mu}.
The best fit to our data corresponds to a scaling law of the form
\begin{gather}
\nonumber
\Delta\mu = a \chi ^{-b} \\
a = 0.88 \pm 0.03 \ \ \  b= 0.65 \pm 0.02 \ 
\end{gather}
with high accuracy as seen in Fig. \ref{fig:delta_mu}.

It is natural to expect the
exponent $b$ to be related to the general critical exponent $\kappa$ introduced in Eq. (\ref{eq:scaling}), since both of them parametrize the effects due to the finiteness of $\chi$.

%%%%%%%%%%%%%%%%%%%%%%%%%%%%%%%%%%%%%%%%%%%%%%%%%%%%%%%%%%%%%%%%%%%%%%%%%%%%%%
%%%%%%%%%%%%%%%%%%%%%%%%%%%%%%%%%%%%%%%%%%%%%%%%%%%%%%%%%%%%%%%%%%%%%%%%%%%%%%
\section{Discussion}

Beyond the well-known numerical strategies for computing a Matrix Product State approximation, exact results for the ground state of a low dimensional MPS can
 be found  using an analytical minimization strategy. The arbitrariness in the parametrization of matrices $\Gamma$ allows us to choose a simple structure for these matrices so that  the minimization problem can be solved analytically. This simplification is guided by the symmetries of the problem.

A number of results emerge from our anaytical approach. First, the exact minimum
found for low $\chi$ cases agrees perfectly with the numerical results obtained
with iTEBD or DMRG. This means that both methods avoid local minima and find
the optimal solution consistent with the MPS ansatz. Second, 
a generic MPS does not preserve any symmetry of the problem beyond the 2-site translational invariance due to the fact that the algorithm looks for a minimum
of the Hamiltonian expectation value. Thus,
for the Heisenberg chain, the 1-site translational invariance of its ground state, and the property of being eigenstate of the total $Z$-component of spin are explicitly violated. However, it is possible to constrain the MPS to preserve one of them while remaining an absolute minimum for the energy. Yet, if both symmetries are imposed at the same time the MPS ansatz does not get down to the same minimum.

For low-$\chi$ MPS it is not difficult to find the suitable parametrization that allows us to preserve a chosen symmetry, as we have shown on examples. It is also possible to 
quantify symmetry restoration using, as a figure of merit, the 
difference of eigenvalues of the 1-body reduced density matrix. We observe
that symmetry gets restored when $\chi$ increases following a detailed scaling law.

The method we have presented can be extended to any local Hamiltonian that displays
translational invariance. This is the key ingredient to reduce the computation to
a simplified minimization. Beyond nearest-neighbor interactions or higher
spins can also be
accommodated with straightforward extensions of the iTEBD algorithm.

\begin{acknowledgements}
We acknowledge fruitful discussions with G. Sierra, S. Iblisdir, 
T. R. de Oliveira and J. M. Escart\'in.
We also ackowledge financial support from MICINN, UE QAP,
SRG.
\end{acknowledgements}

%%%%%%%%%%%%%%%%%%%%%%%%%%%%%%%%%%%%%%%%%%%%%%%%%%%%%%%%%%%%%%%%%%%%%%%%%%%%%%
%%%%%%%%%%%%%%%%%%%%%%%%%%%%%%%%%%%%%%%%%%%%%%%%%%%%%%%%%%%%%%%%%%%%%%%%%%%%%%

\end{document}